# Impacts of Climate Change-Induced Salinity Intrusion on Physiological Parameters of Aquatic Hydrophytes from Coastal Rivers of Bangladesh

Ulfat Jahan Farha [1], Zarin Subah [1], Md Helal Uddin[2], Harunur Rashid [1,2] *

[1] Environmental Sciences Program, Asian University for Women, 20 M.M. Ali Road, Chittagong-4000,
Bangladesh
[2] Department of Fisheries Management, Bangladesh Agricultural University, Mymensingh-2202,
Bangladesh

Email of the corresponding author: harunur7rashid@gmail.com

*Abstract*— Changing temperature, precipitation regimes, and sea level rise - all of these are often associated with a global phenomenon - climate change. An inevitable consequence of this global phenomenon is salinity intrusion which is the gradual movement of salinity into groundwater as well as surface water – mostly through coastal estuaries and rivers. This salinity intrusion can have detrimental impacts on the hydrophytes or the plants living in these water bodies. The current experiment was conducted to see impacts on the physiological traits of some freshwater river hydrophytes – Water Hyacinth (*Eichhornia crassipes*), Buffalo Spinach (*Enhydra fluctuans*), and Taro (*Colocasia esculenta*) against different salinity concentrations (0, 10, 20 and 30 ppt) exposed for 48 hours. The physiological parameters measured were Biomass, Stomata density, Transpiration rate, Total Chlorophyll content, Relative water content, and Histo-architectural changes in roots and stems. A prominent reduction in biomass was noticed with increasing salinity. For stomatal density, it was established that the number of stomata per millimeters square decreased with the increase in salinity concentration. The result for transpiration rate was also in compliance with the result of stomata density. As Taro was observed to be salt tolerant to some extent and had a stomata density of 6 mm $^{-2}$ at 30 ppt, it had the capacity to transpire after 48 hours whereas the rest of the two plants did not have any open stomata at 30 ppt. Hence, there was no transpiration in Water Hyacinth and Buffalo Spinach after 48 hours. Chlorophyll content was also found to be decreasing in Water Hyacinth and Buffalo Spinach, but increasing in Taro with increasing salinity concentration. Relative water content decreased substantially for all the hydrophytes with increasing salinity concentration after 48 hours. Histological study showed deformity in root and tuber tissue structure of Water Hyacinth and Buffalo Spinach. The epidermis of both plants' tissue was thickened at 30 ppt of salt due to the osmotic pressure created by salt stress. There was expansion in the vascular bundle of those tissues which was identified at 30ppt salt which was caused due to the inhibition of the process of water uptake of those plants. All the plants upon prolonged exposure did not survive and the physiological alteration observed and measured after 48 hours justified these conclusions.

## Introduction

Atmospheric concentrations of carbon dioxide have increased by 25% since the Industrial Revolution due to widespread deforestation and fossil fuel combustion (Short and Hilary, 1999). Deforestation and fossil fuel combustion have increased the concentration of carbon dioxide along with other gases such as nitrous oxide, methane, and chlorofluorocarbons (CFCs) which are responsible for absorbing thermal radiation from the surface of the earth and radiating it back to the earth. Due to a significant increase of these radioactive gases, an alarming elevation of Earth's surface temperature has emerged and this constant thermal expansion is predicted to melt the polar ice cap resulting in the expansion of sea level (Short and Hilary, 1999). The current situation of global warming has increased the surface temperature by 2 Fahrenheit which is 1.1° Celsius from the late 19th century and it has greatly been impacted by the increase of greenhouse gases. It is found that the main factor lying behind the huge rise in greenhouse gas emissions is anthropogenic activities (95% of the factors). Increasing livestock farming, fertilizers containing nitrogen produce nitrous oxide emissions, burning coal, oil, and gas, variations in the sun's energy reaching Earth, etc., have accelerated climate change over the years. Climate change, the global threat has been observed to have an effect on large-scale hydrological cycles such as increasing atmospheric water vapor content, changing precipitation patterns and intensity, reduced snow cover, and widespread melting of ice (Bates, 2009; Short and Hilary, 1999). One of the impacts of this drastic change is the increased rate of salinity. Sea level rise due to climate change will cause saline water to migrate to upstream points where only freshwater exists (Pugh, 2004). Researchers have predicted the increase in salinity concentration in estuaries and the change in their circulation due to sea level rise (Short and Hilary 1999). These alterations will eventually result in adverse effects on salt-sensitive habitats. Hydrophytes or plants that grow in water have a moderate range of salinity tolerance beyond which their survival becomes crucial as excessive saline stress impairs their physiological as well as biochemical processes (Haller, 1974; Subah et al., 2018). Research works have confirmed substantial changes in the physiological traits of hydrophytes due to saline stress. For example, salinity reduces $CO_2$ availability in plants which is a result of diffusion limitation which leads to a reduction of the photosynthetic pigments. Machado et al. report that excessive salt concentration leads to salt accumulation which also reduces chlorophyll content and affects the light absorbance by plants (Machado et al., 2017). The growth and survival of hydrophytes are essential for the regulation of native habitats in freshwater. Substantial ecosystem services such as habitat for aquatic animals, nutrient cycling, preservation of water quality, production of wave energy and absorption of wave energy, etc., are dependent upon hydrophytes. Hydrophytes are indispensable not only for their ecological importance but also for their economic significance as they aid in sustaining fisheries as well as water supply. Thus, assessment and quantification of the magnitude to which these hydrophytes can survive and maintain growth are essential along with their physiological and microbial community responses to varying salinity concentrations. Usually, the physiological response of different plants varies due to diverse physiological adaptations (Hu, 2007; Subah et al., 2018). It limits plant growth by disrupting the usual osmosis function of up-taking water and minerals. The salt is dissolved in the water as cation $Na^+$ and anion $Cl^-$ which makes the water toxic for the plants and damages the metabolic system of the plants. Studies of salt-stressed plants are necessary because the development of the germplasms of different species that are salt tolerant can help the survival of different freshwater and terrestrial plants. It will provide food security for a climate change-affected country (Bernstein, 2017; Watson and Lee, 2001). Therefore, this research was conducted to observe and analyze physiological parameters (i.e.,- Biomass yield production, Change in growth, Stomatal density, Transpiration rate, Total chlorophyll content, Relative water content ) of Water Hyacinth (*Eichhornia crassipes*), Helencha (*Enhydra Fluctuans*), Taro (*Colocasia esculenta*) against increasing salinity concentration of 0 ppt, 10 ppt, 20 ppt and 30 ppt.

## METHODS

### Experimental setup:

The plants were collected from different water bodies and were then kept in a setup organized at the rooftop of the Asian University for Women to observe their physiological changes due to different salinity concentrations. The plants that were subjected to salinity were Water Hyacinth *(Eichhornia crassipes)*, Helencha (*Enhydra Fluctuans*), and Taro (*Colocasia esculenta*). Before the plants were settled in the bowls in different salinity concentrations, their roots were gently dried with tissue paper and then the weight and the height of the plants were measured and recorded for further comparison and assessment. The plants were then conserved in concentrations of 0, 10, 20, and 30 ppt of salinity in 6 liters of water mixed with 99% NaCl. Each sample was labeled and replicated thrice in the same concentration for accuracy. Within 48 hours of treatment, the plants were taken into the laboratory for measuring several physiological parameters.

### Biomass of the Plants:

Measuring biomass prior to and subsequent treatment generates an overview of the effect of salinity on biomass. Plants were dried to avoid errors before measurement. Upon measurement of weight using the weighing machine, the plants were kept in treatment for 48 hours and then the mass of the plants was again recorded.

### Height of the plants:

In furtherance of quantifying and assessing the growth of the plant in terms of height, measuring tape was used to determine the height of the plants before and after salinity exposure.

### Stomatal Density :

Pores found in the upper surface of the leaves with two guard cells regulating osmotic changes and transpiration are called stomata. Stomatal density refers to the number of stomata per area. In this experiment in order to measure stomatal density, leaves from each plant from the treatment were collected and the imprint of their upper surface was made. A thin layer of clear nail varnish was used to coat the surface of the leaves and later they were removed with scotch tape by peeling them off. The tape with the leaf impression was then placed on microscope slides (Heidari et al., 2012). Then, the slides were viewed under a microscope at 400x magnification and the radius of the field of view was also measured which was 2 mm. The equation that was followed for measuring stomatal density is described below:

*Number of stomata per field of view/ $\pi r^2$; r = 2 mm*

### Transpiration Rate of the Plants:

Transpiration rate gives an overview of the accurate quantification of the movement of water into plants. In this experiment, 4 similar cuttings with several leaves from each plant were kept in water in 12 measuring cylinders. The water level was adjusted using a teat pipette and the volume of the water was noted. 2 ml of oil was added to the measuring cylinders using a second pipette so that the oil sat on the surface of the water. The water that was lost through the next 72 hours was recorded for each sample in a 24-hour interval period. The transpiration rate per hour was measured using the equation below :

*Transpiration rate = (Final level of water in the cylinder- Initial level of water in the cylinder)/ 24*

### Total Chlorophyll Content:

0.3 g of fresh leaves were taken and ground with 9 ml of 80% acetone in test tubes for each plant from all the treatments. The test tubes were then made airtight and kept in the refrigerator overnight. It was then centrifuged at 5000 rpm for 10 mins. The supernatant was transferred into Eppendorf and the absorbance of the solution was read at 645nm and 663nm against acetone using a spectrophotometer. The formula by Arnon was used for chlorophyll estimation (K. and Banu, 2015). The concentrations of chlorophyll a, chlorophyll b, and total chlorophyll were calculated using the following equation:

*Total Chlorophyll: 20.2(A645) + 8.02(A663)*

### Relative Water Content:

For the relative water content, leaves from each plant were collected and the weight of the leaves was measured using a weighing machine. Fully expanded and mature leaves were selected to measure this parameter and the leaves were cut leaving around 1 cm long petiole. This weight was regarded as the fresh mass of the leaves. Following the measurement of the weight of the leaves the leaves were placed in a ziplock bag. In order to measure the turgor weight of the leaves, 2 ml of 5mM CaCl2 was inserted into the zip bag while the petiole was facing down and only the petiole was in contact with the 5mM CaCl2 solution. The zip lock bag was then closed and the bag was kept in a dark room at room temperature. Eight hours after the preservation of the leaves, they were taken out of the bag and put in between two paper towels in order to lose excess water. The weight of the leaves was measured again which was the turgid weight. Consequently, each leaf sample was inserted into a paper bag to be dried at 60°C for 3 days.

The dried weight of the leaves determined the dry weight and the relative water content of the leaf was measured using the following formula.

*RWC= 100x (fresh mass- dry mass)/ (turgid mass- dry mass)*
(Peñuelas and Inoue, 1999)

***Examination of the Structural Changes of the Root Tissues of Enhydra fluctuans and Tuber Tissues of Eichhornia crassipes:***

The freshwater plants were collected from the place Agrabad Dhebarpara Dighi. The set-up of the experiment was done in the 20J building, M.M Ali Road, Chittagong.
A small portion approximately 5 cm of the tuber of water hyacinth and the root of the Helencha plant was harvested and preserved in FAA solution. The FAA solution was made according to the different proportions of ethanol, glacial acetic acid, formalin, and water. The proportion is written below:
For a 20 ml of FAA solution,
Ethanol (50%)__________________ 10 ml
Glacial Acetic Acid (5%)_________ 1ml
Formalin (10%) _________________ 2ml
Distil Water____________________ 7ml

For preserving the plant tissues of roots and tubers, some 25 ml of small bottles were taken and each bottle was filled up with 20 ml of FAA solution. The plant samples were dipped into the solution and preserved for 4 days at 26 Celsius. After preserving the samples with FAA preservative, the manual sectioning was done. The sectioning was done using a very sharp blade and was stained with Safranin and placed on a glass slide which was mounted with a drop of glycerine. After placing the sectioning in glass slides, coverslips were placed over them and were taken under a microscope for observation. The tissues were observed in 10X ocular and 10X objective lenses and the pictures were taken for further analysis.

## RESULTS

***Biomass Reduction :***

The general trend line in the figure below indicates proportionality between biomass reduction and increased salinity concentration for other plants as well. For example, in terms of *Eichhornia crassip* the mean amount of biomass loss at 20ppt is 11.90 grams, and when it is 30 ppt the biomass reduces to 15.37 grams. (Fig 1)

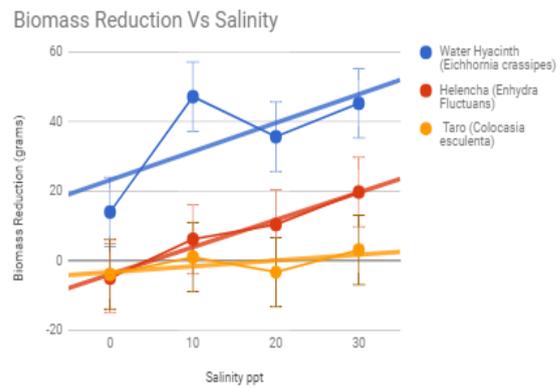

fig .1:Biomass Reduction vs Salinity

***Reduction in Plant Growth:***

As the level of salinity increased the rate of reduction in growth was also observed to be increasing. The graph below indicates the increase in reduction as the concentration of salinity increases. The highest reduction due to increased salinity was observed for the highest salinity concentration of 30 ppt. (Fig 2)

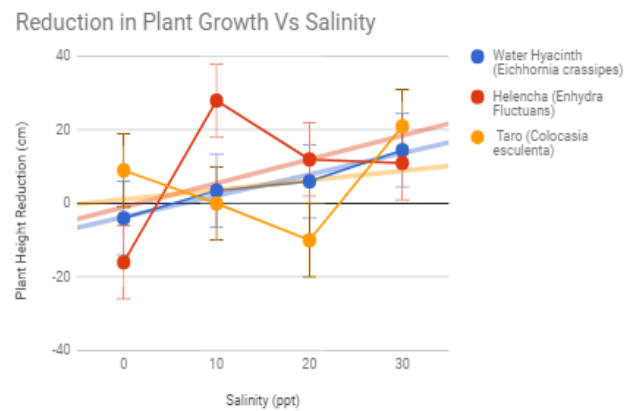

fig 2: Reduction in Plant Growth vs. salinity

### 3.3. Stomatal Density:

Number of open stomata was observed to increase as the salinity concentration went higher. The images below demonstrate the gradual decrease in number of open stomata with increasing higher salinity concentration.

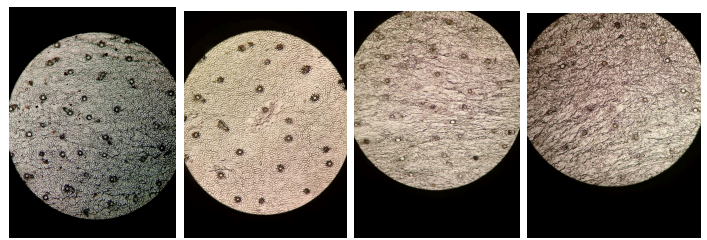

fig 3: Microscopic view of Stomatal density of Enhydra Fluctuans at 400x magnification in 0, 10, 20, and 30 ppt respectively

The general trend refers to the reduction of the number of stomata in 1 mm-2 due to increasing salinity. The highest number of stomata per mm-2 was observed in 0 ppt for Taro and the lowest number of stomata per mm-2 was noticed in Water hyacinth at 30 ppt salinity. (Fig 3)

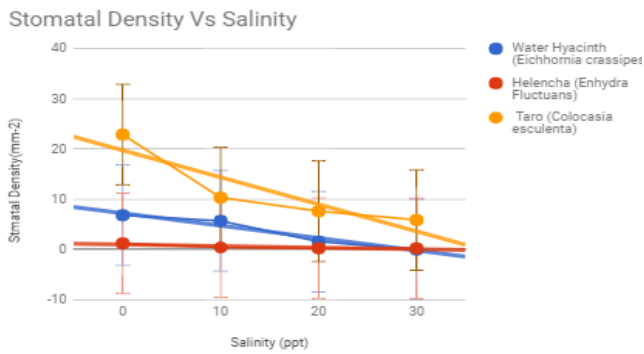

fig 4: Reduction of stomatal density with higher salinity concentration.

### Transpiration Rate:

In this figure, the negative linear correlation expresses the decrease in transpiration rate due to the increase in salinity concentration. For *Colocasia esculenta*, the transpiration rate is relatively higher than for Water Hyacinth and Helencha. The highest transpiration rate was found at .04166 ml/hour in 0 ppt salinity for Taro and Helencha and the lowest rate 0 ml/hr was found in Water hyacinth and Taro. (Figure 4)

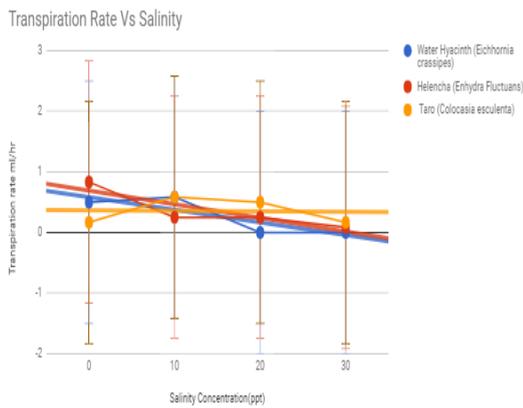

fi 5: Transpiration rate vs. salinity

### Total Chlorophyll Content:

The figure shows the correlation between total chlorophyll content and salinity concentration. The highest average salinity concentration was observed for Taro which was 22.22 μg/ml at 30 ppt and the lowest average concentration was found at 1.11 μg/ml in Water hyacinth at 30 ppt (Figure 5).

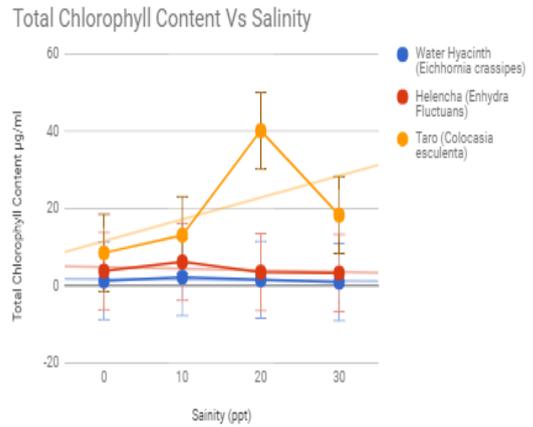

fig 6: Total chlorophyll content vs Salinity

### Relative Water Content:

In this graph, it is evident that RWC decreases significantly with the increase of salinity concentration. The highest average RWC was observed in 0 ppt salinity which was 97.99% in Taro and the lowest average RWC was observed in Taro which was 37.48%. (Figure 6)

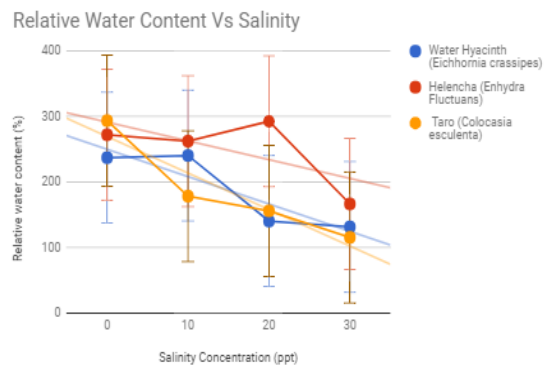

fig 7: Relative Water Content vs. salinity Concentration

### Examination of the Structural Changes of the Root Tissues of Enhydra fluctuans and Tuber Tissues of Eichhornia crassipes:

### Root Tissues of Enhydra fluctuans in 0ppt:

The microscopic photo of the tissue (Figure 8) showed a clear view of the outer layer of the epidermis, cortex containing parenchyma cells, then sclerenchyma cells or blast fiber, and the darker part in the middle clearly showed vascular tissues xylem and phloem. The clear view of the cross-section of *Enhydra fluctuans* in 0ppt is shown below:

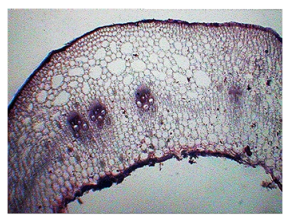

fig 8: Enhydra fluctuans tissue in 0ppt

### Root Tissues of *Enhydra fluctuans* in 30ppt:

The deformity in the cell structure is seen in the thick layer of the epidermis and the dense structure of the cortex (Figure 9). Though there are not any specific changes detected in the xylem cells besides the diameter of xylem cells being reduced, the phloem cells are increased in number. The collenchyma is seen to be thickened in the wall and densely surrounding the parenchyma cells. The microscopic view shows the differences in the tissue of 30ppt from the control one. The differences are marked in the pics.

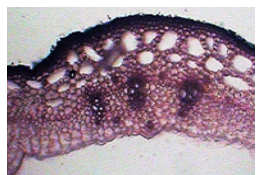

*fig 9: Enhydra fluctuans Root Tissue in 30 ppt, showing the differences in the thick epidermis, cell density, and vascular bundle expansion*

### Tuber Tissues of *Eichhornia crassipes* in 0 ppt:

Under the microscope, the tissue showed a clear view of the outer layer of the epidermis, cortex containing aerenchyma, and parenchyma cells(Fig 10 (a,b). The aerenchyma is one of the hydromorphic traits that was found in the tissue of *Eichhornia crassipes* in 0ppt. This aerenchyma is also called air spaces which were clearly identified under the microscope. The darker part in the middle showed vascular bundles including the xylem and phloem which were hard to identify due to not good sectioning. The clear view of the cross-section of *Eichhornia crassipes* in 0ppt is shown below:

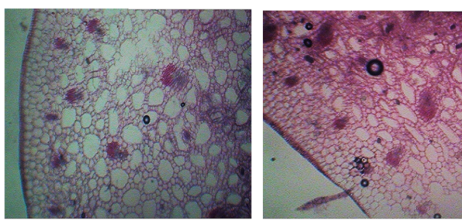

*fig 10: Water Hyacinth Tuber Tissue 0ppt*

### Tuber Tissues of *Eichhornia crassipes* in 30 ppt:

The distinctive differences can be seen in the tuber tissue of *Eichhornia crassipes* which was exposed in the 30ppt of saline water. The changes in the cell structure showed the disruption in the epidermis cells (Fig 11 (a,b)). The thick layer of epidermis and dense structure of the cortex were identified in the tissues. The aerenchyma cells were found squeezed and not in the specific structures. But there were not any distinct changes detected in the vascular bundle, besides the diameter of the vascular bundle being increased in number. The outer cortex is seen to be thickened in the cell wall and densely surrounding the parenchyma cells. The microscopic view clearly shows the differences in the tissue of 30 ppt from the control one. The differences are marked in the photos.

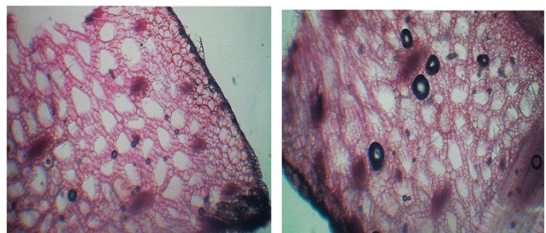

*fig 11: Water Hyacinth Tuber Tissue 30 ppt, differences are showing in the epidermis and the vascular bundle of the tissue*

## DISCUSSION

The observed plants showed significant alteration in their physiological characteristics according to different levels of salinity. This result has been in compliance with other researchers' conclusions who have described the hyperosmotic and hyperionic as the reason for plant death due to high salinity. The impacts on different parameters have been described below.

### Effect on Biomass:

In terms of biomass, it was observed that upon exposure, the biomass for the plants decreased in treatment plants rather than in control which was 0 ppt. By observing the weight of the plants under salt stress after the treatment interval, it was noticed that a general trend appeared for the decrease in weight of the plants against increasing salinity concentration. The graph 1.1 indicates a positive linear relationship between biomass reduction and salinity concentration. The results found in this study is in agreement with the results found in an experiment by Ambede et al., where the decrease in plant biomass was recorded with the increase in salinity concentration *Vigna subterranea (L.) Verdc (*Ambede et al., 2012). The highest reduction was seen in 30 ppt for Water Hyacinth and Helencha with a mean reduction of 15.11 grams and 6.6 grams respectively. This reduction can be a plausible result of a combination of specific ion of Na+ and Cl- and increased osmotic pressure. The high salinity concentration reduced the water acquisition capability of plants which induced metabolic changes to the treatment plants. Increased osmotic pressure induces reduction in cytoplasmic volume which initiates the loss of cell turgor. Other research works have also observed opposite results where biomass increased with increased salinity (Alam et al., 2015). In *Colocasia esculenta* the reduction in growth was lowest and the average mass reduction that occurred in this plant at highest salinity stress was 1.03 grams. This is an indication of Taro's high salt tolerance capacity than the two other plants. Other research works have also supported Taro being salt tolerant to permissible level.

### Effect on Plant Growth:

The effect of increased salinity concentration can be seen from the linear positive correlation from graph 2.1. For all three plants Water Hyacinth (*Eichhornia crassipes*),

Helencha (*Enhydra Fluctuans*) and Taro (*Colocasia esculenta*) the reduction in height was increased in terms of increasing salinity concentration which is in agreement with other research works that have been conducted in this regard. From graph 2.1 it can be seen that the rate of loss in reduction was highest in 30 ppt which is the highest salinity concentration. The lowest loss in reduction was observed in Taro as it has been mentioned earlier that this plant has salt tolerant capacity among the three plants and thus has been affected less in terms of the two other plants. Plant growth has been reported to be affected by salinity due to nutritional imbalances as well as specific ion toxicities. This reduction has been described to happen in two stages by Munns et al. The first stage includes water deficit-induced rapid reduction of growth and the second stage is ascribed to a gradual accumulation of salts in the shoot of the plants at a toxic level (Machado et al., 2017). This reduction due to salinity concentration has been observed by Bernardo et al., as mentioned earlier.

### *Effect on Stomatal Density:*

A decrease in the number of open stomata has been observed in the plants as the concentration of salinity went higher. The linear negative correlation illustrated by the 3.1 graph depicts that the stomatal density decreased for all plants as the salinity concentration increased. Taro showed the highest decrease in terms of the number of open stomata and Water Hyacinth showed the lowest number of stomata in 30 ppt among the three plants. The value for stomatal density was 6/mm$^2$ in Taro whereas for Water hyacinth and Helencha, there was no open stomata at 30 ppt which indicates that transpiration stopped for Water hyacinth and Helencha at 30 ppt. As the level of salinity goes high the functionality of the plant's metabolism decreases hence the reduction of stomatal density is justified. Under saline stress plants' inability to acquire essential amounts of water leads to closure of stomata so that the plants do not lose excessive water. Moreover, the plants uptake sodium and chloride ions, which interfere with the uptake of potassium ions disrupt stomatal regulation, and cause necrosis. This phenomenon of stomatal closure due to high salinity has also been reported by other research works (Heidari et al., 2012). This stomatal closure is attributed to reduced photosynthesis and reduced intercellular carbon dioxide concentration.

### *Effect on Transpiration Rate:*

As the stomatal density decreases with increasing salinity level, the rate of transpiration is supposed to decline and the observed data concur with the theory. It was found that at 30 ppt rate of transpiration becomes very low in Taro at .083 ml/hr whereas in the case of *Enhydra Fluctuans* and of *Eichhornia crassipes* the transpiration rate was 0 ml/hr (Mert and Vardar, 1977). This result is in parallel with the result found for stomatal density. As no open stomata were found in the plants at 30 ppt, it was expected that at this salinity concentration, there would not be any transpiration and the result was found to be legitimate. Uptake of injurious ions by the plants leads to accumulation of these ions in plants leading to damaging chloroplasts, protein synthesis as well as other organelles. These disruptions impact the transpiration rate of the older leaves of the plants (Machado et al., 2017; Nuffieldfoundation.org., 2017) Another impact of this toxic ion accumulation is deficiency of other essential nutrients. For example, it has been found that a high concentration of Na+ in water initiates calcium deficiency in plants and this has led to a decreased transpiration rate in tomato plants (Machado et al., 2017). Experimentation by Jamil and Rha has also observed a remarkable reduction in transpiration rate in mustard due to sodium chloride stress (Jamil et al., 2013).

### *Effect on Total Chlorophyll Content*

Potential photosynthetic productivity and plant vigor can be indicated by chlorophyll content as the essential components required for photosynthesis exist in chloroplast at specific molar ratios to chlorophyll. The figure 5.1 shows that with the increasing salinity concentration, total chlorophyll content in Water Hyacinth and Helencha decreases. The highest average concentration for salinity was observed at 22.22 μg/ml at 30 ppt and the lowest average concentration was found at 1.11 μg/ml in Water hyacinth at 30 ppt. For Taro, the general trend line shows a positive relation between total chlorophyll content and salinity concentration. A plausible explanation for this result could be that in salt-tolerant plants, chlorophyll content increases with accession to salinity concentration (Heidari et al., 2012) Similarly as taro has salt tolerance capacity it exhibits this positive relation. In Water Hyacinth and Helencha, the linear relationship is negative as saline stress aggravates enzymatic chlorophyll degradation. Salinity induces the weakening of the protein-pigment-lipid complex which increases chlorophyllase activity; that is the initial step of the breakdown of chlorophyll. Another factor that could contribute to the reduction of chlorophyll content is the reduction of leaf area due to high salinity concentration that led to lower light interception. Heidari et al. observed similar consequences for increased salinity concentration in *Ocimum basilicum L.*

### *Effect on Relative Water Content*

It can be observed from Graph 6.1 that the relative water content of the plants was adversely affected due to increased salinity. This reduction in RWC is the result of turgor loss due to limited water availability induced by salinity. RWC estimates the existent water content of the sample leaf in terms of the ratio of the maximum amount of water the leaf can hold at full turgidity. Usually, the normal values of RWC range between ninety-eight percent in turgid leaves and forty percent in severely desiccated leaves (Heidari et al., 2012). At the highest salinity, the average RWC was found to be 37.48% which is much lesser. A significant decrease was observed in Taro relative to water hyacinth and Helencha. That indicates the other two plants' adaptability in order to minimize water loss under saline stress. Another investigation in this regard has also assured rapid reduction of RWC due to high salinity in *Triticum aestivum L* (El-Bassiouny et al., 2005).

### *Examination of the Structural Changes of the Root Tissues of Enhydra fluctuans and Tuber Tissues of Eichhornia crassipes:*

The differences were found in the freshwater plant tissue due to salinity increase in those plants. The common histological change that was found in both *Eichhornia crassipes* and *Enhydra fluctuans* was the thickening of the epidermis cells. The introduction of salinity caused the reduction of certain necessary minerals of $Ca^+$ and $K^+$ which led the root cells to be limited from the deficit minerals. Also, osmotic pressure increased in the cells of the roots of the plants which prohibited the water uptake from the soil. Thus the unfavorable condition was created due to salinity intrusion in plants, the cell wanted to prevent it through the defense mechanism of building a thicker epidermis layer and not letting salt enter the tissue. Similarly, an experiment done by Shabala also describes the extension of the epidermis for the salinity effect (Shabala, 2006).

Besides the epidermis extension, the diameter of the vascular bundle also increased due to salt stress. Both plants *Eichhornia crassipes* and *Enhydra fluctuans* showed the expansion of the vascular bundle in the salt stress. The xylem expansion was seen in his experiment which can happen because of lignifications and scarcity of water to uptake (Sánchez-Aguayo et al., 2004). Moreover, another experiment which was done by Kiegle et al., on the response of cells to salt in the *Arabidopsis* root showed the result that salt stress created huge differences in the cell structure. The salt concentration that was provided to the plants was 220mM NaCl. The cells of the root of the plants showed prolonged endodermis and pericycle. This prolonged structure was distinct from the other control plant's cells.

The same reason can be given behind the expansion of the xylem of vascular bundle in 30 ppt of *Eichhornia crassipes* and *Enhydra fluctuans*. The water shortage due to the osmotic pressure created because salt water increases, has led the changes in the structure of the vascular bundle tissues (Imchen et al., 2017). Thus there were distinct changes in the epidermis and vascular bundle of the freshwater were detected in this study due to high salinity (Sheldon et al., 2017). The effect of osmotic pressure and defense mechanisms in the cells of those plants resulted in their histo-architecture changes. There were some resource and time limitation factors present in this study for which the histology technique with dehydration, embedding, sectioning with microtome, and standard staining processes couldn't be followed. The limitation in this study was one of the factors which was the constraint in identifying the changes on a very cellular basis. Thus this study needs to be continued further by using new and recent techniques of histology to identify the precise changes in the salinity-affected plant tissues.

## Conclusion

Global climate change has become one of the major problems of this century which has compelled various problems in the environment i,e, global temperature rise, ocean acidification, sea level rise, shrinking ice sheets, salinity intrusion, etc. A large source of primary production is dependent upon the survivability of the aquatic ecosystems. Salinity intrusion will hamper the range of salinity in freshwater resources which will eventually impact the distribution of the flora in the aquatic habitat. Although in this experiment Taro had relatively less impact among the three plants due to its salt tolerance capacity, with prolonged exposure it could not survive as the increased salinity acted at a toxic level for the hydrophytes. In terms of the observation of the histology, the epidermis of *Eichhornia crassipes* and *Enhydra fluctuans* tissues was thickened and the vascular bundle was expanded for the introduction of the saline water in the plants. A distinct type of defense mechanism was developed in the plant tissue which acted against the salt intrusion in the tissues. With the limited range of salt tolerance, the hydrophytes will not be able to survive and the consequences of its destruction will rebound to the humans who initiated the process of climate change in the first place. Proper measurements to decelerate global climate change have to be implemented to ensure the impregnability of not only the aquatic ecosystem but also the terrestrial ecosystem that may be susceptible to the consequences of global climate change.